\documentclass[aps,prb,twocolumn,showpacs,preprintnumbers,superscriptaddress]{revtex4}
\usepackage{graphicx}
\usepackage{dcolumn}
\begin{document}

\title{Calorimetric Evidence of Multiband Superconductivity in Ba(Fe$_{0.925}$Co$_{0.075}$)$_{2}$As$_{2}$}
\author{F. Hardy}
\email[]{Frederic.Hardy@ifp.fzk.de}
\author{T. Wolf}
\affiliation{Karlsruher Institut f\"ur Technologie, Institut f\"ur Festk\"orperphysik, 76021 Karlsruhe, Germany}
\author{R. A. Fisher}
\affiliation{Lawrence Berkeley National Laboratory, Berkeley CA 94720, USA}
\author{R. Eder}
\author{P. Schweiss}
\author{P. Adelmann}
\affiliation{Karlsruher Institut f\"ur Technologie, Institut f\"ur Festk\"orperphysik, 76021 Karlsruhe, Germany}
\author{H. v. L\"ohneysen}
\affiliation{Karlsruher Institut f\"ur Technologie, Institut f\"ur Festk\"orperphysik, 76021 Karlsruhe, Germany}
\affiliation{Karlsruher Institut f\"ur Technologie, Physikalisches Institut, 76128 Karlsruhe, Germany}
\author{C. Meingast}
\affiliation{Karlsruher Institut f\"ur Technologie, Institut f\"ur Festk\"orperphysik, 76021 Karlsruhe, Germany}
\date{\today}
\begin{abstract}
We report on the determination of the electronic heat capacity of a slightly overdoped (x = 0.075) Ba(Fe$_{1-x}$Co$_{x}$)$_{2}$As$_{2}$ single crystal with a T$_{c}$ of 21.4 K. Our analysis of the temperature dependence of the superconducting-state specific heat provides strong evidence for a two-band {\it s}-wave order parameter with gap amplitudes 2$\Delta_{1}$(0)/k$_{B}$T$_{c}$=1.9 and 2$\Delta_{2}$(0)/k$_{B}$T$_{c}$=4.4.  Our result is consistent with the recently predicted $s_{+-}$ order parameter [I. I. Mazin et al., {\it Phys. Rev. Lett.} 101, 057003 (2008)].
\end{abstract}
\maketitle


%

The newly discovered iron arsenide family (FeAs) offers new possibilities for studying the interplay between superconductivity and magnetism.~\cite{Kamihara08,Rotter08,Sefat08} As for many other materials, {\it e.g.}, heavy fermions and cuprates, superconductivity emerges in the vicinity of a magnetic instability. The origin of the pairing interaction, as well as the gap symmetry remain unidentified in the pnictides. Theoretically, the particular topology of the Fermi surface with strong nesting features favor a multiband order parameter having either an $s_{+-}$-wave or a $d$-wave symmetry.~\cite{Mazin08,Kuroki08,Bang08} In either case, a $\pi$-shift of the order-parameter phase is expected between different sheets of the Fermi surface. The identification of the gap symmetry is crucial, because it will shed light on the mechanism responsible for the condensation of Cooper pairs. Experimentally, solid evidence for a particular pairing state remains elusive, because several experimental probes point to different conclusions. For instance, in the electron-doped 122 compound, Ba(Fe$_{1-x}$Co$_{x}$)$_{2}$As$_{2}$, photoemission data~\cite{Terashima09} (ARPES), and point-contact spectroscopy~\cite{Samuely09} show two distinct nodeless gaps with large amplitudes, while penetration-depth measurements~\cite{Gordon09} exhibit a power-law behavior reflecting the possible existence of nodes. Similar discrepancies are observed for hole-doped (Ba$_{1-x}$K$_{x}$)Fe$_{2}$As$_{2}$ and the 1111 series.~\cite{Mu09,Martin09} Some of these apparent contradictions may arise from the influence of the magnetic instability, which is expected to strongly alter the gap topology,~\cite{Parker09} from impurity effects, or from experimental difficulties like sample inhomogeneities or surface off-stoichiometry. Specific-heat measurements can provide an important measure of the bulk superconducting properties; specifically, they can give valuable information about the possible existence of nodes in the energy gap and, as previously shown for MgB$_{2}$,~\cite{Fisher03,Bouquet01} to the number of bands that contribute to the superconducting condensate. Several specific-heat measurements have been reported for the Fe-pnictides, but the interpretation of the results has been impaired by substantial contributions from paramagnetic centers and/or an incorrect evaluation of the large phonon background.~\cite{Mu09,Welp09,Ding08,Budko09}

In this Letter, we present a detailed analysis of the electronic specific heat of a slightly overdoped Ba(Fe$_{1-x}$Co$_{x}$)$_{2}$As$_{2}$ single crystal with x=0.075, {\it i.e.} at a doping level where the static spin-density wave (SDW) is no longer observed.  The problem of the phonon background determination is overcome by measuring a strongly overdoped crystal with x=0.153, in which superconductivity is suppressed. Our analysis for the superconducting sample (x=0.075) gives strong evidence for two energy gaps, which implies that several sheets of the Fermi surface contribute to the formation of Cooper pairs. Additionally, we provide reliable values of the normal-state Sommerfeld coefficients $\gamma_{n}$=C$_{e}$/T for several Co concentrations.
\begin{figure}[h]
\begin{center}
\includegraphics[width=8.5cm]{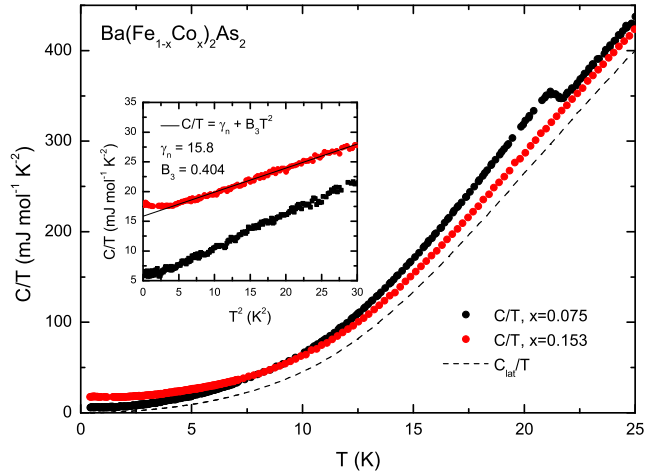}
\caption{\label{fig:Fig1} Temperature dependence of the specific heat C of samples with x=0.075 and x=0.153. The dashed line represents the lattice contribution, C$_{lat}$/T, derived from the specific heat for x=0.153 (see text). The inset shows the low-temperature specific heat of both samples.}
\end{center}
\end{figure}

Co-doped Ba122 single crystals were grown from self-flux in glassy carbon crucibles. Prereacted FeAs and CoAs powders were mixed with Ba, placed into the crucible, which then was sealed in an evacuated SiO$_{2}$ ampoule. After heating to 650 $^{\circ}$C and then to $\approx$ 1200 $^{\circ}$C with holding times of 5 hours, crystal growth took place during cooling at a rate of $\approx$ 1 $^{\circ}$C/h. At 1000 $^{\circ}$C, the ampoule was tilted to decant the remaining liquid flux from the crystals and subsequently removed from the furnace. The composition of these samples was determined by energy dispersive x-ray spectroscopy to be x = 0.075 ($\pm$ 0.005-0.01) and x = 0.153 ($\pm$ 0.005-0.01), respectively. The specific heat was measured with the $^{3}$He option in a PPMS from Quantum Design.\\

Figure \ref{fig:Fig1} shows that the specific heat of both samples is dominated by the lattice contribution; the electronic part is only about  $\approx$ 10 \% of the total signal at T$_{c}$. Therefore, it is impossible to obtain an accurate and unique description of the lattice background down to T = 0 K using the harmonic-lattice approximation, {\it i.e.} by fitting the specific heat to an odd-power polynomial in a restricted range above T$_{c}$. As will be demonstrated below, a much more reliable phonon specific heat, C$_{lat}$ (dashed line in Fig.\ref{fig:Fig1}) is obtained using the specific heat of the x=0.153 sample (for T$>$2 K) after subtraction of a constant electronic term $\gamma_{n}$= 15.8 mJ mol$^{-1}$ K$^{-2}$. The electronic term of this sample is obtained by fitting the data in the inset of Fig.\ref{fig:Fig1} with electronic and phononic terms, while ignoring the small peak at $\approx$ 0.7 K, which may be due either to traces of remaining superconductivity or to the contribution from paramagnetic centers.  It should be pointed out that the data of the superconducting sample (x=0.075) contain a significant residual linear term of about 6 mJ mol$^{-1}$ K$^{-2}$, which will be discussed later.  No traces of the long range SDW were detected down to 0.5 K for either sample, in agreement with previous reports.~\cite{Chu09,Ni08,Budko09} 
\begin{figure}[h]
\begin{center}
\includegraphics[width=8.5cm]{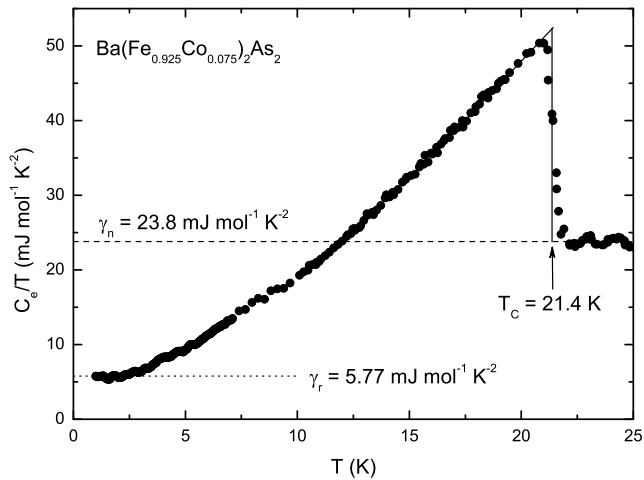}
\caption{\label{fig:Fig2} Temperature dependence of the electron specific heat, C$_{e}$/T, of the superconducting sample (x=0.075). The dashed line represents
the normal-state electron contribution, $\gamma_{n}$=23.8 mJ mol$^{-1}$ K$^{-2}$. The dotted line is a residual normal-state-like contribution, $\gamma_{r}$=5.77 mJ mol$^{-1}$ K$^{-2}$.}
\end{center}
\end{figure}

Figure \ref{fig:Fig2} shows the temperature dependence of the electron specific heat C$_{e}$ (x=0.075) = C(x=0.075) - $f_{s}$$\cdot$ C$_{lat}$(x=0.153), with a scaling factor $f_{s}$ of 1.01. The factor $f_{s}$ is introduced because it is not expected that the phonon specific heats of both samples are strictly identical. The magnitude of $f_{s}$ was determined by enforcing entropy conservation, {\it i.e.} $\int_{0}^{T_{c}} \gamma_{n} dT$ = $\int_{0}^{T_{c}} C_{e}/T dT$. The small deviation of $f_{s}$ from unity demonstrates that the above procedure represents a very good method to determine the phonon background.  Physically, this can be attributed to the fact that the substitution of Fe by Co does not substantially affect the lattice properties, as shown by recent inelastic x-ray scattering measurements and ab-initio calculations.~\cite{Reznik09} 

\begin{figure}[h]
\begin{center}
\includegraphics[width=8.5cm]{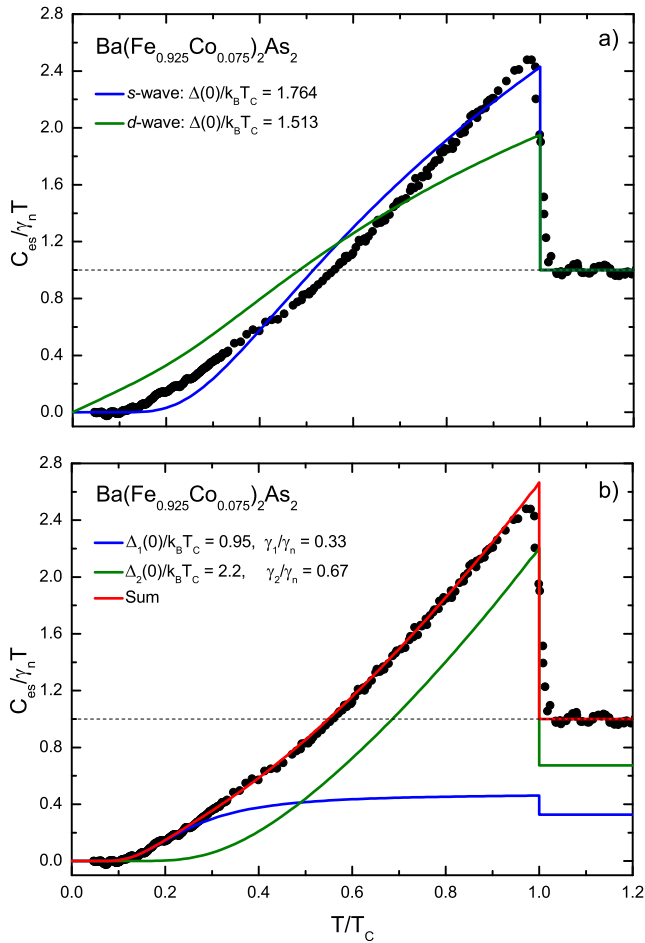}
\caption{\label{fig:Fig3} a) The electron specific heat of the superconducting sample (x=0.075), normalized to 1 mol of superconducting condensate, compared with the specific heat of single-band {\it s}-wave (blue line) and {\it d}-wave (green line) order parameters, in the weak coupling limit.
b) The electron specific heat of the superconducting sample (x=0.075), normalized to 1 mole of superconducting condensate. The red curve represents a two-gap fit. The blue and green curves are the partial specific-heat contributions of the two bands.}
\end{center}
\end{figure}

\begin{table}[h]
\caption{\label{tab:Table1} Critical temperature (T$_{c}$) and normal-state electron specific heat ($\gamma_{n}$). Value for x=0 is taken from Ref.~\cite{fred-un}}
\begin{ruledtabular}
\begin{tabular}{c|cc}
x&T$_{c}$(K)&$\gamma_{n}$(mJ mol$^{-1}$ K$^{-2}$)\\ 
\hline
0					&			0.0			&		5.3\\
0.075			&     21.4		&		23.8\\
0.153			&			0.7 		&		15.8\\
\end{tabular}
\end{ruledtabular}
\end{table}
The superconducting transition at T$_{c}$=21.4 K is remarkably sharp, indicating little inhomogeneity in the crystal. The normal-state electron contribution for x=0.075 is $\gamma_{n}$ $\approx$ 24 mJ mol$^{-1}$ K$^{-2}$, which is in excellent agreement with LDA+DMFT calculations, which require $\gamma_{n}$ to be enhanced to 20-30 mJ mol$^{-1}$ K$^{-2}$ in order to explain mass renormalization by Hund coupling.~\cite{Haule09} Our values for several Co concentrations, summarized in Table \ref{tab:Table1}, show that the disappearance of the SDW with Co-doping is accompanied by an increase of the electronic density of states (EDOS), compatible with a progressive closure of the SDW gap. In the overdoped region, on the other hand, $\gamma_{n}$ and T$_{c}$ both decrease. Interestingly, $\gamma_{n}$ of our superconducting sample is only about half as large as the value reported for K-doped 122 single crystals~\cite{Mu09} ($\approx$ 63 mJ mol$^{-1}$ K$^{-2}$).  Figure \ref{fig:Fig2} illustrates that C$_{e}$/T does not extrapolate to zero at T=0 but to a residual normal-state-like contribution $\gamma_{r}$=5.8 mJ mol$^{-1}$ K$^{-2}$.  Taken at face value, this would indicate that the sample has a superconducting fraction of $\approx$ 75 \%. Finite values of $\gamma_{r}$ are a general feature of specific-heat measurements of electron-doped 122 iron arsenides.~\cite{Mu09,Mu09-2} For the cuprates, they have been attributed to an incomplete transition to the superconducting-state and volume fractions of normal and superconducting material $\gamma_{r}$/$\gamma_{n}$ and 1 - $\gamma_{r}$/$\gamma_{n}$, respectively. On this basis, the specific heat is the sum of separate contributions of the superconducting and normal phases and consequently, the electronic specific heat can be normalized to one mole of superconducting material, C$_{es}$, defined by:
\begin{equation}\label{eq:eq1}
C_{es}=(C_{e}-\gamma_{r}T)\cdot\frac{\gamma_{n}}{\gamma_{n}-\gamma_{r}}
\end{equation}
However, recent specific-heat~\cite{Mu09-2} and heat-transport~\cite{Machida09} measurements suggest that $\gamma_{r}$T is a consequence of pair breaking in electron-doped 122 pnictides, and not due to the presence of normal material. Thus, if the $s_{+-}$ state is present,~\cite{Mazin08} $\gamma_{r}$T can be understood as arising from interband scattering, induced by in-plane disorder due to Co doping, which is pair-breaking for a sign-reversing order parameter.~\cite{Onari09} In that case, the specific heat is, in principle, not the sum of contributions of broken pairs and the superconducting condensate. Nevertheless, in analogy with the Na cobaltates,~\cite{Oeschler08} C$_{es}$ (given by Eq.\ref{eq:eq1}) can be expected to be a reasonable and useful approximation to the specific heat of one mole of superconducting material, and is used, in Fig. \ref{fig:Fig3}, for the purpose of comparison with several possible order parameters.
Figure \ref{fig:Fig3}(a) demonstrates that C$_{es}$ cannot be described by the specific heat of a single-band BCS {\it s}-wave superconductor, calculated in the weak-coupling limit (blue line). The agreement is very poor. As for MgB$_{2}$,~\cite{Fisher03} the positive curvature of C$_{es}$ for T/T$_{c}$$>$0.6, where the BCS curve shows negative curvature, is indicative of strong coupling effects, but the observed discontinuity at T$_{c}$, $\Delta$C$_{es}$/$\gamma_{n}$T$_{c}$, which would be greater than the BCS value for a strong-coupled single-band superconductor, is close to weak-coupling value. In addition, C$_{es}$ is significantly larger than the BCS curve, for T/T$_{c}$$<$0.4, again arguing against strong-coupling effects. Figure \ref{fig:Fig3}(a) also shows the specific heat of a single-band {\it d}-wave superconductor in the weak coupling limit (green line).\footnote{We assumed a cylindrical Fermi surface. For this geometry, $d_{x^{2}-y^{2}}$ and $d_{xy}$ order parameters exhibit the same temperature-dependent specific heat.} It is obvious that such a {\it k}-dependent gap, even in a strong-coupling or a two-band scenario, cannot describe the observed low-temperature exponential behavior that can be inferred from the data.

We therefore focus our discussion on the possibility of two energy gaps, using the phenomenological two-band $\alpha$-model, introduced by Bouquet {\it et al}.~\cite{Bouquet01,Padamsee73} It allows a fit of the specific heat from low temperatures up to T$_{c}$ and, as a result, gives reliable gap amplitudes that were shown to agree quantitatively with band calculations on MgB$_{2}$ in particular,~\cite{Bouquet01,Liu01} and with Eliashberg equations in general.~\cite{Dolgov05} In this fit, the specific heat is taken as the sum of contributions from two bands, which are calculated independently assuming a BCS temperature dependence of the superconducting gaps. Two gap magnitudes, at T=0 , are introduced as adjustable parameters, $\alpha_{1}$=$\Delta_{1}$(0)/k$_{B}$T$_{c}$ and $\alpha_{2}$=$\Delta_{2}$(0)/k$_{B}$T$_{c}$, together with a third quantity, $\gamma_{i}$/$\gamma_{n}$ ({\it i}=1, 2), which measures the fraction of the total normal EDOS that the {\it i}-th band contributes to the superconducting condensate.\footnote{with the constraint that $\gamma_{1}$/$\gamma_{n}$+$\gamma_{2}$/$\gamma_{n}$=1} As shown in Fig.\ref{fig:Fig3}(b), the two-band fit (with $\alpha_{1}$=0.95, $\alpha_{2}$=2.2
and $\gamma_{1}$/$\gamma_{n}$=0.33) accurately reproduces the specific heat over the entire temperature range. The smaller gap is about half the weak-coupling BCS value $\Delta_{BCS}(0)$=1.764, while the second gap is larger. These values are comparable with those derived from recent NMR (in the large scattering-rate limit~\cite{Wang09}) and $\mu$SR-penetration-depth measurements~\cite{Williams09} (see Table \ref{tab:Table2}), but differ appreciably, by at least a factor 1.5, from ARPES data.~\cite{Terashima09}
\begin{table}[h]
\caption{\label{tab:Table2} Gap ratios 2$\Delta_{1}$(0)/k$_{B}$T$_{c}$, 2$\Delta_{2}$(0)/k$_{B}$T$_{c}$ and weights $\gamma_{1}$/$\gamma_{n}$ as determined by the two-gap model and by different techniques.~\cite{Terashima09,Wang09,Williams09}.}
\begin{ruledtabular}
\begin{tabular}{c|c|c|c|c}
Technique			&	x								& 2$\Delta_{1}$(0)/k$_{B}$T$_{c}$			& 2$\Delta_{2}$(0)/k$_{B}$T$_{c}$						&  $\gamma_{1}$/$\gamma_{n}$\\ 
\hline
C (T)					&	0.075						&	1.9																	&4.4																				&  0.33\\
NMR						&	0.070						&	1.8																	&7.2																				&  0.4\\
$\mu$SR				&	0.070						&	1.565																&3.768																			&  0.345\\
ARPES					&	0.075						&	4.1																	&6.4																				&  -\\
\end{tabular}
\end{ruledtabular}
\end{table}

The temperature dependence of the superconducting-state specific heat, as well as the substantial residual EDOS, is consistent with the predicted extended {\it s}-wave order parameter.~\cite{Mazin08,Kuroki08,Bang08} However, the major gap develops around the Fermi-surface sheet that shows the largest EDOS while it is theoretically expected that $\Delta_{1}/\Delta_{2}\propto\sqrt{N_{2}/N_{1}}$, in the limit of pure interband pairing.~\cite{Dolgov09} Thus, our results indicate that intraband interactions are more important than expected. Ba(Fe$_{0.925}$Co$_{0.075}$)$_{2}$As$_{2}$ also shares similar properties with NbSe$_{2}$, another candidate for multiband superconductivity, as illustrated by thermodynamic measurements~\cite{Huang07,Boaknin03} and ARPES spectra.~\cite{Yokoya01} The specific heat of both NbSe$_{2}$ and Ba(Fe$_{0.925}$Co$_{0.075}$)$_{2}$As$_{2}$ show no sign of an incipient steep increase of C(T) below T$_{c}$, which is the conspicuous signature of the small gap in MgB$_{2}$.~\cite{Fisher03} This difference can be understood in terms of the gap anisotropy, $\Delta_{2}$/$\Delta_{1}$, and the EDOS ratio, $\gamma_{2}$/$\gamma_{1}$. This steep increase is particularly pronounced in MgB$_{2}$ because (i) $\Delta_{2}$/$\Delta_{1}$ is about twice as large as in Ba(Fe$_{0.925}$Co$_{0.075}$)$_{2}$As$_{2}$ and NbSe$_{2}$, (ii) each gap gives an equal contribution, $\gamma_{2}$/$\gamma_{1}$$\approx$1, to the specific heat of MgB$_{2}$. In contrast, in both NbSe$_{2}$ and Ba(Fe$_{0.925}$Co$_{0.075}$)$_{2}$As$_{2}$ the major-gap contributions strongly dominate, with $\gamma_{2}$/$\gamma_{1}$ roughly equal to 4 and 2.3, respectively.

In summary, a detailed analysis of the electronic specific heat of Ba(Fe$_{0.925}$Co$_{0.075}$)$_{2}$As$_{2}$ provides strong evidence of a multigap order parameter, as observed for MgB$_{2}$ and NbSe$_{2}$.  Our data are fit very well by a two-band {\it s}-wave model, and our results are, thus, in agreement with the predicted $s_{+-}$ pairing-state.  Further, we derive a reliable phonon contribution that permits to extract accurate values of the normal-state Sommerfeld coefficients. 

\bibliography{biblio}

\begin{thebibliography}{35}
\expandafter\ifx\csname natexlab\endcsname\relax\def\natexlab#1{#1}\fi
\expandafter\ifx\csname bibnamefont\endcsname\relax
  \def\bibnamefont#1{#1}\fi
\expandafter\ifx\csname bibfnamefont\endcsname\relax
  \def\bibfnamefont#1{#1}\fi
\expandafter\ifx\csname citenamefont\endcsname\relax
  \def\citenamefont#1{#1}\fi
\expandafter\ifx\csname url\endcsname\relax
  \def\url#1{\texttt{#1}}\fi
\expandafter\ifx\csname urlprefix\endcsname\relax\def\urlprefix{URL }\fi
\providecommand{\bibinfo}[2]{#2}
\providecommand{\eprint}[2][]{\url{#2}}

\bibitem[{\citenamefont{Kamihara et~al.}(2008)\citenamefont{Kamihara, Watanabe,
  Hirano, and Hosono}}]{Kamihara08}
\bibinfo{author}{\bibfnamefont{Y.}~\bibnamefont{Kamihara}},
  \bibinfo{author}{\bibfnamefont{T.}~\bibnamefont{Watanabe}},
  \bibinfo{author}{\bibfnamefont{M.}~\bibnamefont{Hirano}}, \bibnamefont{and}
  \bibinfo{author}{\bibfnamefont{H.}~\bibnamefont{Hosono}},
  \bibinfo{journal}{J.\ Am.\ Chem.\ Soc.} \textbf{\bibinfo{volume}{130}},
  \bibinfo{pages}{3296} (\bibinfo{year}{2008}).

\bibitem[{\citenamefont{Rotter et~al.}(2008)\citenamefont{Rotter, Pangerl,
  Tegel, and Johrendt}}]{Rotter08}
\bibinfo{author}{\bibfnamefont{M.}~\bibnamefont{Rotter}},
  \bibinfo{author}{\bibfnamefont{M.}~\bibnamefont{Pangerl}},
  \bibinfo{author}{\bibfnamefont{M.}~\bibnamefont{Tegel}}, \bibnamefont{and}
  \bibinfo{author}{\bibfnamefont{D.}~\bibnamefont{Johrendt}},
  \bibinfo{journal}{Angew.\ Chem.\ Int.\ Ed.} \textbf{\bibinfo{volume}{47}},
  \bibinfo{pages}{7949} (\bibinfo{year}{2008}).

\bibitem[{\citenamefont{Sefat et~al.}(2008)\citenamefont{Sefat, Jin, McGuire,
  Sales, Singh, and Mandrus}}]{Sefat08}
\bibinfo{author}{\bibfnamefont{A.~S.} \bibnamefont{Sefat}},
  \bibinfo{author}{\bibfnamefont{R.}~\bibnamefont{Jin}},
  \bibinfo{author}{\bibfnamefont{M.~A.} \bibnamefont{McGuire}},
  \bibinfo{author}{\bibfnamefont{B.~C.} \bibnamefont{Sales}},
  \bibinfo{author}{\bibfnamefont{D.~J.} \bibnamefont{Singh}}, \bibnamefont{and}
  \bibinfo{author}{\bibfnamefont{D.}~\bibnamefont{Mandrus}},
  \bibinfo{journal}{Phys.\ Rev.\ Lett.} \textbf{\bibinfo{volume}{101}},
  \bibinfo{pages}{117004} (\bibinfo{year}{2008}).

\bibitem[{\citenamefont{Mazin et~al.}(2008)\citenamefont{Mazin, Singh,
  Johannes, and Du}}]{Mazin08}
\bibinfo{author}{\bibfnamefont{I.~I.} \bibnamefont{Mazin}},
  \bibinfo{author}{\bibfnamefont{D.~J.} \bibnamefont{Singh}},
  \bibinfo{author}{\bibfnamefont{M.~D.} \bibnamefont{Johannes}},
  \bibnamefont{and} \bibinfo{author}{\bibfnamefont{M.~H.} \bibnamefont{Du}},
  \bibinfo{journal}{Phys.\ Rev.\ Lett.} \textbf{\bibinfo{volume}{101}},
  \bibinfo{pages}{057003} (\bibinfo{year}{2008}).

\bibitem[{\citenamefont{Kuroki et~al.}(2008)\citenamefont{Kuroki, Onari, Arita,
  Usui, Tanaka, Kontani, and Aoki}}]{Kuroki08}
\bibinfo{author}{\bibfnamefont{K.}~\bibnamefont{Kuroki}},
  \bibinfo{author}{\bibfnamefont{S.}~\bibnamefont{Onari}},
  \bibinfo{author}{\bibfnamefont{R.}~\bibnamefont{Arita}},
  \bibinfo{author}{\bibfnamefont{H.}~\bibnamefont{Usui}},
  \bibinfo{author}{\bibfnamefont{Y.}~\bibnamefont{Tanaka}},
  \bibinfo{author}{\bibfnamefont{H.}~\bibnamefont{Kontani}}, \bibnamefont{and}
  \bibinfo{author}{\bibfnamefont{H.}~\bibnamefont{Aoki}},
  \bibinfo{journal}{Phys.\ Rev.\ Lett.} \textbf{\bibinfo{volume}{101}},
  \bibinfo{pages}{087004} (\bibinfo{year}{2008}).

\bibitem[{\citenamefont{Bang and Choi}(2008)}]{Bang08}
\bibinfo{author}{\bibfnamefont{Y.}~\bibnamefont{Bang}} \bibnamefont{and}
  \bibinfo{author}{\bibfnamefont{H.-Y.} \bibnamefont{Choi}},
  \bibinfo{journal}{Phys.\ Rev.\ B} \textbf{\bibinfo{volume}{78}},
  \bibinfo{pages}{134523} (\bibinfo{year}{2008}).

\bibitem[{\citenamefont{Terashima et~al.}(2009)\citenamefont{Terashima, Sekiba,
  Bowen, Nakayama, Kawahara, Sato, Richard, Xu, Li, Cao et~al.}}]{Terashima09}
\bibinfo{author}{\bibfnamefont{K.}~\bibnamefont{Terashima}},
  \bibinfo{author}{\bibfnamefont{Y.}~\bibnamefont{Sekiba}},
  \bibinfo{author}{\bibfnamefont{J.~H.} \bibnamefont{Bowen}},
  \bibinfo{author}{\bibfnamefont{K.}~\bibnamefont{Nakayama}},
  \bibinfo{author}{\bibfnamefont{T.}~\bibnamefont{Kawahara}},
  \bibinfo{author}{\bibfnamefont{T.}~\bibnamefont{Sato}},
  \bibinfo{author}{\bibfnamefont{P.}~\bibnamefont{Richard}},
  \bibinfo{author}{\bibfnamefont{Y.~M.} \bibnamefont{Xu}},
  \bibinfo{author}{\bibfnamefont{L.~J.} \bibnamefont{Li}},
  \bibinfo{author}{\bibfnamefont{G.~H.} \bibnamefont{Cao}},
  \bibnamefont{et~al.}, \bibinfo{journal}{Proc.\ Natl.\ Acad.\ Sci.\ USA}
  \textbf{\bibinfo{volume}{106}}, \bibinfo{pages}{7330} (\bibinfo{year}{2009}).

\bibitem[{\citenamefont{Samuely et~al.}(2009)\citenamefont{Samuely,
  Pribulov\'a, Szab\'o, Prist\'a\v{s}, Bud'ko, and Canfield}}]{Samuely09}
\bibinfo{author}{\bibfnamefont{P.}~\bibnamefont{Samuely}},
  \bibinfo{author}{\bibfnamefont{Z.}~\bibnamefont{Pribulov\'a}},
  \bibinfo{author}{\bibfnamefont{P.}~\bibnamefont{Szab\'o}},
  \bibinfo{author}{\bibfnamefont{G.}~\bibnamefont{Prist\'a\v{s}}},
  \bibinfo{author}{\bibfnamefont{S.~L.} \bibnamefont{Bud'ko}},
  \bibnamefont{and} \bibinfo{author}{\bibfnamefont{P.~C.}
  \bibnamefont{Canfield}}, \bibinfo{journal}{Physica \ C}
  \textbf{\bibinfo{volume}{469}}, \bibinfo{pages}{507} (\bibinfo{year}{2009}).

\bibitem[{\citenamefont{Gordon et~al.}(2009)\citenamefont{Gordon, Martin, Kim,
  Ni, Tanatar, Schmalian, Mazin, Bud'ko, Canfield, and Prozorov}}]{Gordon09}
\bibinfo{author}{\bibfnamefont{R.~T.} \bibnamefont{Gordon}},
  \bibinfo{author}{\bibfnamefont{C.}~\bibnamefont{Martin}},
  \bibinfo{author}{\bibfnamefont{H.}~\bibnamefont{Kim}},
  \bibinfo{author}{\bibfnamefont{N.}~\bibnamefont{Ni}},
  \bibinfo{author}{\bibfnamefont{M.~A.} \bibnamefont{Tanatar}},
  \bibinfo{author}{\bibfnamefont{J.}~\bibnamefont{Schmalian}},
  \bibinfo{author}{\bibfnamefont{I.~I.} \bibnamefont{Mazin}},
  \bibinfo{author}{\bibfnamefont{S.~L.} \bibnamefont{Bud'ko}},
  \bibinfo{author}{\bibfnamefont{P.~C.} \bibnamefont{Canfield}},
  \bibnamefont{and} \bibinfo{author}{\bibfnamefont{R.}~\bibnamefont{Prozorov}},
  \bibinfo{journal}{Phys.\ Rev.\ B} \textbf{\bibinfo{volume}{79}},
  \bibinfo{pages}{100506} (\bibinfo{year}{2009}).

\bibitem[{\citenamefont{Mu et~al.}(2009{\natexlab{a}})\citenamefont{Mu, Luo,
  Wang, Shan, Ren, and Wen}}]{Mu09}
\bibinfo{author}{\bibfnamefont{G.}~\bibnamefont{Mu}},
  \bibinfo{author}{\bibfnamefont{H.}~\bibnamefont{Luo}},
  \bibinfo{author}{\bibfnamefont{Z.}~\bibnamefont{Wang}},
  \bibinfo{author}{\bibfnamefont{L.}~\bibnamefont{Shan}},
  \bibinfo{author}{\bibfnamefont{C.}~\bibnamefont{Ren}}, \bibnamefont{and}
  \bibinfo{author}{\bibfnamefont{H.-H.} \bibnamefont{Wen}},
  \bibinfo{journal}{Phys.\ Rev.\ B} \textbf{\bibinfo{volume}{79}},
  \bibinfo{pages}{174501} (\bibinfo{year}{2009}{\natexlab{a}}).

\bibitem[{\citenamefont{Martin et~al.}(2009)\citenamefont{Martin, Gordon,
  Tanatar, Kim, Ni, Bud'ko, Canfield, Luo, Wen, Wang et~al.}}]{Martin09}
\bibinfo{author}{\bibfnamefont{C.}~\bibnamefont{Martin}},
  \bibinfo{author}{\bibfnamefont{R.~T.} \bibnamefont{Gordon}},
  \bibinfo{author}{\bibfnamefont{M.~A.} \bibnamefont{Tanatar}},
  \bibinfo{author}{\bibfnamefont{H.}~\bibnamefont{Kim}},
  \bibinfo{author}{\bibfnamefont{N.}~\bibnamefont{Ni}},
  \bibinfo{author}{\bibfnamefont{S.~L.} \bibnamefont{Bud'ko}},
  \bibinfo{author}{\bibfnamefont{P.~C.} \bibnamefont{Canfield}},
  \bibinfo{author}{\bibfnamefont{H.}~\bibnamefont{Luo}},
  \bibinfo{author}{\bibfnamefont{H.~H.} \bibnamefont{Wen}},
  \bibinfo{author}{\bibfnamefont{Z.}~\bibnamefont{Wang}}, \bibnamefont{et~al.},
  \bibinfo{journal}{Phys.\ Rev.\ B.} \textbf{\bibinfo{volume}{80}},
  \bibinfo{pages}{020501} (\bibinfo{year}{2009}).

\bibitem[{\citenamefont{Parker et~al.}(2009)\citenamefont{Parker, Vavilov,
  Chubukov, and Mazin}}]{Parker09}
\bibinfo{author}{\bibfnamefont{D.}~\bibnamefont{Parker}},
  \bibinfo{author}{\bibfnamefont{M.~G.} \bibnamefont{Vavilov}},
  \bibinfo{author}{\bibfnamefont{A.~V.} \bibnamefont{Chubukov}},
  \bibnamefont{and} \bibinfo{author}{\bibfnamefont{I.~I.} \bibnamefont{Mazin}},
  \bibinfo{journal}{Phys.\ Rev.\ B} \textbf{\bibinfo{volume}{80}},
  \bibinfo{pages}{100508} (\bibinfo{year}{2009}).

\bibitem[{\citenamefont{Fisher et~al.}(2003)\citenamefont{Fisher, Li, Lashley,
  Bouquet, Phillips, Hinks, Jorgensen, and Crabtree}}]{Fisher03}
\bibinfo{author}{\bibfnamefont{R.~A.} \bibnamefont{Fisher}},
  \bibinfo{author}{\bibfnamefont{G.}~\bibnamefont{Li}},
  \bibinfo{author}{\bibfnamefont{J.~C.} \bibnamefont{Lashley}},
  \bibinfo{author}{\bibfnamefont{F.}~\bibnamefont{Bouquet}},
  \bibinfo{author}{\bibfnamefont{N.~E.} \bibnamefont{Phillips}},
  \bibinfo{author}{\bibfnamefont{D.~G.} \bibnamefont{Hinks}},
  \bibinfo{author}{\bibfnamefont{J.~D.} \bibnamefont{Jorgensen}},
  \bibnamefont{and} \bibinfo{author}{\bibfnamefont{G.~W.}
  \bibnamefont{Crabtree}}, \bibinfo{journal}{Physica \ C}
  \textbf{\bibinfo{volume}{385}}, \bibinfo{pages}{180} (\bibinfo{year}{2003}).

\bibitem[{\citenamefont{Bouquet et~al.}(2001)\citenamefont{Bouquet, Wang,
  Fisher, Hinks, Jorgensen, Junod, and Phillips}}]{Bouquet01}
\bibinfo{author}{\bibfnamefont{F.}~\bibnamefont{Bouquet}},
  \bibinfo{author}{\bibfnamefont{Y.}~\bibnamefont{Wang}},
  \bibinfo{author}{\bibfnamefont{R.~A.} \bibnamefont{Fisher}},
  \bibinfo{author}{\bibfnamefont{D.~G.} \bibnamefont{Hinks}},
  \bibinfo{author}{\bibfnamefont{J.~D.} \bibnamefont{Jorgensen}},
  \bibinfo{author}{\bibfnamefont{A.}~\bibnamefont{Junod}}, \bibnamefont{and}
  \bibinfo{author}{\bibfnamefont{N.~E.} \bibnamefont{Phillips}},
  \bibinfo{journal}{Europhys.\ Lett.} \textbf{\bibinfo{volume}{56}},
  \bibinfo{pages}{856} (\bibinfo{year}{2001}).

\bibitem[{\citenamefont{Welp et~al.}(2009)\citenamefont{Welp, Mu, Xie,
  Koshelev, Kwok, Luo, Wang, Cheng, Fang, Ren et~al.}}]{Welp09}
\bibinfo{author}{\bibfnamefont{U.}~\bibnamefont{Welp}},
  \bibinfo{author}{\bibfnamefont{G.}~\bibnamefont{Mu}},
  \bibinfo{author}{\bibfnamefont{R.}~\bibnamefont{Xie}},
  \bibinfo{author}{\bibfnamefont{A.~E.} \bibnamefont{Koshelev}},
  \bibinfo{author}{\bibfnamefont{W.~K.} \bibnamefont{Kwok}},
  \bibinfo{author}{\bibfnamefont{H.~Q.} \bibnamefont{Luo}},
  \bibinfo{author}{\bibfnamefont{Z.~S.} \bibnamefont{Wang}},
  \bibinfo{author}{\bibfnamefont{P.}~\bibnamefont{Cheng}},
  \bibinfo{author}{\bibfnamefont{L.}~\bibnamefont{Fang}},
  \bibinfo{author}{\bibfnamefont{C.}~\bibnamefont{Ren}}, \bibnamefont{et~al.},
  \bibinfo{journal}{Physica \ C} \textbf{\bibinfo{volume}{469}},
  \bibinfo{pages}{575} (\bibinfo{year}{2009}).

\bibitem[{\citenamefont{Ding et~al.}(2008)\citenamefont{Ding, He, Dong, Wu,
  Liu, Chen, and Li}}]{Ding08}
\bibinfo{author}{\bibfnamefont{L.}~\bibnamefont{Ding}},
  \bibinfo{author}{\bibfnamefont{C.}~\bibnamefont{He}},
  \bibinfo{author}{\bibfnamefont{J.~K.} \bibnamefont{Dong}},
  \bibinfo{author}{\bibfnamefont{T.}~\bibnamefont{Wu}},
  \bibinfo{author}{\bibfnamefont{R.~H.} \bibnamefont{Liu}},
  \bibinfo{author}{\bibfnamefont{X.~H.} \bibnamefont{Chen}}, \bibnamefont{and}
  \bibinfo{author}{\bibfnamefont{S.~Y.} \bibnamefont{Li}},
  \bibinfo{journal}{Phys.\ Rev.\ B} \textbf{\bibinfo{volume}{77}},
  \bibinfo{pages}{180510} (\bibinfo{year}{2008}).

\bibitem[{\citenamefont{Bud'ko et~al.}(2009)\citenamefont{Bud'ko, Ni, and
  Canfield}}]{Budko09}
\bibinfo{author}{\bibfnamefont{S.~L.} \bibnamefont{Bud'ko}},
  \bibinfo{author}{\bibfnamefont{N.}~\bibnamefont{Ni}}, \bibnamefont{and}
  \bibinfo{author}{\bibfnamefont{P.~C.} \bibnamefont{Canfield}},
  \bibinfo{journal}{Phys.\ Rev.\ B} \textbf{\bibinfo{volume}{79}},
  \bibinfo{pages}{220516(R)} (\bibinfo{year}{2009}).

\bibitem[{\citenamefont{Chu et~al.}(2009)\citenamefont{Chu, Analytis,
  Kucharczyk, and Fisher}}]{Chu09}
\bibinfo{author}{\bibfnamefont{J.-H.} \bibnamefont{Chu}},
  \bibinfo{author}{\bibfnamefont{J.~G.} \bibnamefont{Analytis}},
  \bibinfo{author}{\bibfnamefont{C.}~\bibnamefont{Kucharczyk}},
  \bibnamefont{and} \bibinfo{author}{\bibfnamefont{I.~R.}
  \bibnamefont{Fisher}}, \bibinfo{journal}{Phys.\ Rev.\ B.}
  \textbf{\bibinfo{volume}{79}}, \bibinfo{pages}{187004}
  (\bibinfo{year}{2009}).

\bibitem[{\citenamefont{Ni et~al.}(2008)\citenamefont{Ni, Tillman, Yan,
  Kracher, Hannahs, Bud'ko, and Canfield}}]{Ni08}
\bibinfo{author}{\bibfnamefont{N.}~\bibnamefont{Ni}},
  \bibinfo{author}{\bibfnamefont{M.~E.} \bibnamefont{Tillman}},
  \bibinfo{author}{\bibfnamefont{J.-Q.} \bibnamefont{Yan}},
  \bibinfo{author}{\bibfnamefont{A.}~\bibnamefont{Kracher}},
  \bibinfo{author}{\bibfnamefont{S.-T.} \bibnamefont{Hannahs}},
  \bibinfo{author}{\bibfnamefont{S.~L.} \bibnamefont{Bud'ko}},
  \bibnamefont{and} \bibinfo{author}{\bibfnamefont{P.~C.}
  \bibnamefont{Canfield}}, \bibinfo{journal}{Phys.\ Rev.\ B}
  \textbf{\bibinfo{volume}{78}}, \bibinfo{pages}{214515}
  (\bibinfo{year}{2008}).

\bibitem[{\citenamefont{Reznik et~al.}(2009)\citenamefont{Reznik, Lokshin,
  Mitchell, Parshall, Dmowski, Lamago, Heid, Bohnen, Sefat, McGuire
  et~al.}}]{Reznik09}
\bibinfo{author}{\bibfnamefont{D.}~\bibnamefont{Reznik}},
  \bibinfo{author}{\bibfnamefont{K.}~\bibnamefont{Lokshin}},
  \bibinfo{author}{\bibfnamefont{D.~C.} \bibnamefont{Mitchell}},
  \bibinfo{author}{\bibfnamefont{D.}~\bibnamefont{Parshall}},
  \bibinfo{author}{\bibfnamefont{W.}~\bibnamefont{Dmowski}},
  \bibinfo{author}{\bibfnamefont{D.}~\bibnamefont{Lamago}},
  \bibinfo{author}{\bibfnamefont{R.}~\bibnamefont{Heid}},
  \bibinfo{author}{\bibfnamefont{K.~P.} \bibnamefont{Bohnen}},
  \bibinfo{author}{\bibfnamefont{A.~S.} \bibnamefont{Sefat}},
  \bibinfo{author}{\bibfnamefont{M.~A.} \bibnamefont{McGuire}},
  \bibnamefont{et~al.}, \bibinfo{journal}{arXiv:0810.4941v1}
  (\bibinfo{year}{2009}).

\bibitem[{\citenamefont{Hardy}(2009)}]{fred-un}
\bibinfo{author}{\bibfnamefont{F.}~\bibnamefont{Hardy}},
  \bibinfo{journal}{unpublished}  (\bibinfo{year}{2009}).

\bibitem[{\citenamefont{Haule and Kotliar}(2009)}]{Haule09}
\bibinfo{author}{\bibfnamefont{K.}~\bibnamefont{Haule}} \bibnamefont{and}
  \bibinfo{author}{\bibfnamefont{G.}~\bibnamefont{Kotliar}},
  \bibinfo{journal}{Proceedings of the International Conference on Magnetism,
  Karlsruhe}  (\bibinfo{year}{2009}).

\bibitem[{\citenamefont{Mu et~al.}(2009{\natexlab{b}})\citenamefont{Mu, Zheng,
  Cheng, Wang, Fang, Shen, Shan, Ren, and Wen}}]{Mu09-2}
\bibinfo{author}{\bibfnamefont{G.}~\bibnamefont{Mu}},
  \bibinfo{author}{\bibfnamefont{B.}~\bibnamefont{Zheng}},
  \bibinfo{author}{\bibfnamefont{P.}~\bibnamefont{Cheng}},
  \bibinfo{author}{\bibfnamefont{Z.}~\bibnamefont{Wang}},
  \bibinfo{author}{\bibfnamefont{L.}~\bibnamefont{Fang}},
  \bibinfo{author}{\bibfnamefont{B.}~\bibnamefont{Shen}},
  \bibinfo{author}{\bibfnamefont{L.}~\bibnamefont{Shan}},
  \bibinfo{author}{\bibfnamefont{C.}~\bibnamefont{Ren}}, \bibnamefont{and}
  \bibinfo{author}{\bibfnamefont{H.~H.} \bibnamefont{Wen}},
  \bibinfo{journal}{arXiv:0906.4513v2}  (\bibinfo{year}{2009}{\natexlab{b}}).

\bibitem[{\citenamefont{Machida et~al.}(2009)\citenamefont{Machida, Tomokuni,
  Isono, Izawa, Nakajima, and Tamegai}}]{Machida09}
\bibinfo{author}{\bibfnamefont{Y.}~\bibnamefont{Machida}},
  \bibinfo{author}{\bibfnamefont{K.}~\bibnamefont{Tomokuni}},
  \bibinfo{author}{\bibfnamefont{T.}~\bibnamefont{Isono}},
  \bibinfo{author}{\bibfnamefont{K.}~\bibnamefont{Izawa}},
  \bibinfo{author}{\bibfnamefont{Y.}~\bibnamefont{Nakajima}}, \bibnamefont{and}
  \bibinfo{author}{\bibfnamefont{T.}~\bibnamefont{Tamegai}},
  \bibinfo{journal}{J.\ Phys.\ Soc.\ Japan} \textbf{\bibinfo{volume}{78}},
  \bibinfo{pages}{073705} (\bibinfo{year}{2009}).

\bibitem[{\citenamefont{Onari and Kontani}(2009)}]{Onari09}
\bibinfo{author}{\bibfnamefont{S.}~\bibnamefont{Onari}} \bibnamefont{and}
  \bibinfo{author}{\bibfnamefont{H.}~\bibnamefont{Kontani}},
  \bibinfo{journal}{arXiv:0906.2269v1}  (\bibinfo{year}{2009}).

\bibitem[{\citenamefont{Oeschler et~al.}(2008)\citenamefont{Oeschler, Fisher,
  Phillips, Gordon, Foo, and Cava}}]{Oeschler08}
\bibinfo{author}{\bibfnamefont{N.}~\bibnamefont{Oeschler}},
  \bibinfo{author}{\bibfnamefont{R.~A.} \bibnamefont{Fisher}},
  \bibinfo{author}{\bibfnamefont{N.~E.} \bibnamefont{Phillips}},
  \bibinfo{author}{\bibfnamefont{J.~E.} \bibnamefont{Gordon}},
  \bibinfo{author}{\bibfnamefont{M.-F.} \bibnamefont{Foo}}, \bibnamefont{and}
  \bibinfo{author}{\bibfnamefont{R.~J.} \bibnamefont{Cava}},
  \bibinfo{journal}{Phys.\ Rev.\ B.} \textbf{\bibinfo{volume}{78}},
  \bibinfo{pages}{054528} (\bibinfo{year}{2008}).

\bibitem[{\citenamefont{Padamsee et~al.}(1973)\citenamefont{Padamsee, Neighbor,
  and Shiffman}}]{Padamsee73}
\bibinfo{author}{\bibfnamefont{H.}~\bibnamefont{Padamsee}},
  \bibinfo{author}{\bibfnamefont{J.~E.} \bibnamefont{Neighbor}},
  \bibnamefont{and} \bibinfo{author}{\bibfnamefont{C.~A.}
  \bibnamefont{Shiffman}}, \bibinfo{journal}{J.\ Low\ Temp.\ Phys.}
  \textbf{\bibinfo{volume}{12}}, \bibinfo{pages}{387} (\bibinfo{year}{1973}).

\bibitem[{\citenamefont{Liu et~al.}(2001)\citenamefont{Liu, Mazin, and
  Kortus}}]{Liu01}
\bibinfo{author}{\bibfnamefont{A.}~\bibnamefont{Liu}},
  \bibinfo{author}{\bibfnamefont{I.~I.} \bibnamefont{Mazin}}, \bibnamefont{and}
  \bibinfo{author}{\bibfnamefont{J.}~\bibnamefont{Kortus}},
  \bibinfo{journal}{Phys.\ Rev.\ Lett.} \textbf{\bibinfo{volume}{87}},
  \bibinfo{pages}{087005} (\bibinfo{year}{2001}).

\bibitem[{\citenamefont{Dolgov et~al.}(2005)\citenamefont{Dolgov, Kremer,
  Kortus, Golubov, and Shulga}}]{Dolgov05}
\bibinfo{author}{\bibfnamefont{O.~V.} \bibnamefont{Dolgov}},
  \bibinfo{author}{\bibfnamefont{R.~K.} \bibnamefont{Kremer}},
  \bibinfo{author}{\bibfnamefont{J.}~\bibnamefont{Kortus}},
  \bibinfo{author}{\bibfnamefont{A.~A.} \bibnamefont{Golubov}},
  \bibnamefont{and} \bibinfo{author}{\bibfnamefont{S.~V.}
  \bibnamefont{Shulga}}, \bibinfo{journal}{Phys.\ Rev.\ B}
  \textbf{\bibinfo{volume}{72}}, \bibinfo{pages}{024504}
  (\bibinfo{year}{2005}).

\bibitem[{\citenamefont{Q.Zheng et~al.}(2009)\citenamefont{Q.Zheng, Matano,
  Kawasaki, Ren, Zhao, Chen, Luo, Wang, and Lin}}]{Wang09}
\bibinfo{author}{\bibfnamefont{G.}~\bibnamefont{Q.Zheng}},
  \bibinfo{author}{\bibfnamefont{K.}~\bibnamefont{Matano}},
  \bibinfo{author}{\bibfnamefont{S.}~\bibnamefont{Kawasaki}},
  \bibinfo{author}{\bibfnamefont{Z.~A.} \bibnamefont{Ren}},
  \bibinfo{author}{\bibfnamefont{Z.~X.} \bibnamefont{Zhao}},
  \bibinfo{author}{\bibfnamefont{G.~F.} \bibnamefont{Chen}},
  \bibinfo{author}{\bibfnamefont{J.~L.} \bibnamefont{Luo}},
  \bibinfo{author}{\bibfnamefont{N.~L.} \bibnamefont{Wang}}, \bibnamefont{and}
  \bibinfo{author}{\bibfnamefont{C.~T.} \bibnamefont{Lin}},
  \bibinfo{journal}{Proceedings of the International Conference on Magnetism,
  Karlsruhe}  (\bibinfo{year}{2009}).

\bibitem[{\citenamefont{Williams et~al.}(2009)\citenamefont{Williams, Aczel,
  Baggio-Saitovitch, Bud'ko, Canfield, Carlo, Goko, Munevar, Ni, Uemura
  et~al.}}]{Williams09}
\bibinfo{author}{\bibfnamefont{T.~J.} \bibnamefont{Williams}},
  \bibinfo{author}{\bibfnamefont{A.~A.} \bibnamefont{Aczel}},
  \bibinfo{author}{\bibfnamefont{E.}~\bibnamefont{Baggio-Saitovitch}},
  \bibinfo{author}{\bibfnamefont{S.~L.} \bibnamefont{Bud'ko}},
  \bibinfo{author}{\bibfnamefont{P.~C.} \bibnamefont{Canfield}},
  \bibinfo{author}{\bibfnamefont{J.~P.} \bibnamefont{Carlo}},
  \bibinfo{author}{\bibfnamefont{T.}~\bibnamefont{Goko}},
  \bibinfo{author}{\bibfnamefont{J.}~\bibnamefont{Munevar}},
  \bibinfo{author}{\bibfnamefont{N.}~\bibnamefont{Ni}},
  \bibinfo{author}{\bibfnamefont{Y.~J.} \bibnamefont{Uemura}},
  \bibnamefont{et~al.}, \bibinfo{journal}{Phys.\ Rev.\ B}
  \textbf{\bibinfo{volume}{80}}, \bibinfo{pages}{094501}
  (\bibinfo{year}{2009}).

\bibitem[{\citenamefont{Dolgov et~al.}(2009)\citenamefont{Dolgov, Mazin,
  Parker, and Golubov}}]{Dolgov09}
\bibinfo{author}{\bibfnamefont{O.~V.} \bibnamefont{Dolgov}},
  \bibinfo{author}{\bibfnamefont{I.~I.} \bibnamefont{Mazin}},
  \bibinfo{author}{\bibfnamefont{D.}~\bibnamefont{Parker}}, \bibnamefont{and}
  \bibinfo{author}{\bibfnamefont{A.~A.} \bibnamefont{Golubov}},
  \bibinfo{journal}{Phys.\ Rev.\ B} \textbf{\bibinfo{volume}{79}},
  \bibinfo{pages}{060502} (\bibinfo{year}{2009}).

\bibitem[{\citenamefont{Huang et~al.}(2007)\citenamefont{Huang, Lin, Chang,
  Sun, Shen, Chou, Berger, Lee, and Yang}}]{Huang07}
\bibinfo{author}{\bibfnamefont{C.~L.} \bibnamefont{Huang}},
  \bibinfo{author}{\bibfnamefont{J.-Y.} \bibnamefont{Lin}},
  \bibinfo{author}{\bibfnamefont{Y.~T.} \bibnamefont{Chang}},
  \bibinfo{author}{\bibfnamefont{C.~P.} \bibnamefont{Sun}},
  \bibinfo{author}{\bibfnamefont{H.~Y.} \bibnamefont{Shen}},
  \bibinfo{author}{\bibfnamefont{C.~C.} \bibnamefont{Chou}},
  \bibinfo{author}{\bibfnamefont{H.}~\bibnamefont{Berger}},
  \bibinfo{author}{\bibfnamefont{T.~K.} \bibnamefont{Lee}}, \bibnamefont{and}
  \bibinfo{author}{\bibfnamefont{H.~D.} \bibnamefont{Yang}},
  \bibinfo{journal}{Phys.\ Rev.\ B} \textbf{\bibinfo{volume}{76}},
  \bibinfo{pages}{212504} (\bibinfo{year}{2007}).

\bibitem[{\citenamefont{Boaknin et~al.}(2003)\citenamefont{Boaknin, Tanatar,
  Paglione, Hawthorn, Ronning, Hill, Sutherland, Taillefer, Sonier, Hayden
  et~al.}}]{Boaknin03}
\bibinfo{author}{\bibfnamefont{E.}~\bibnamefont{Boaknin}},
  \bibinfo{author}{\bibfnamefont{M.~A.} \bibnamefont{Tanatar}},
  \bibinfo{author}{\bibfnamefont{J.}~\bibnamefont{Paglione}},
  \bibinfo{author}{\bibfnamefont{D.}~\bibnamefont{Hawthorn}},
  \bibinfo{author}{\bibfnamefont{F.}~\bibnamefont{Ronning}},
  \bibinfo{author}{\bibfnamefont{R.~W.} \bibnamefont{Hill}},
  \bibinfo{author}{\bibfnamefont{M.}~\bibnamefont{Sutherland}},
  \bibinfo{author}{\bibfnamefont{L.}~\bibnamefont{Taillefer}},
  \bibinfo{author}{\bibfnamefont{J.}~\bibnamefont{Sonier}},
  \bibinfo{author}{\bibfnamefont{S.~M.} \bibnamefont{Hayden}},
  \bibnamefont{et~al.}, \bibinfo{journal}{Phys.\ Rev.\ Lett.}
  \textbf{\bibinfo{volume}{90}}, \bibinfo{pages}{117003}
  (\bibinfo{year}{2003}).

\bibitem[{\citenamefont{Yokoya et~al.}(2001)\citenamefont{Yokoya, Kiss,
  Chainani, Shin, Nohara, and Takagi}}]{Yokoya01}
\bibinfo{author}{\bibfnamefont{T.}~\bibnamefont{Yokoya}},
  \bibinfo{author}{\bibfnamefont{T.}~\bibnamefont{Kiss}},
  \bibinfo{author}{\bibfnamefont{A.}~\bibnamefont{Chainani}},
  \bibinfo{author}{\bibfnamefont{S.}~\bibnamefont{Shin}},
  \bibinfo{author}{\bibfnamefont{M.}~\bibnamefont{Nohara}}, \bibnamefont{and}
  \bibinfo{author}{\bibfnamefont{H.}~\bibnamefont{Takagi}},
  \bibinfo{journal}{Science} \textbf{\bibinfo{volume}{294}},
  \bibinfo{pages}{2518} (\bibinfo{year}{2001}).

\end{thebibliography}
\end{document}